\documentclass{Interspeech2024}
\usepackage{tikz}
\usepackage{pgfplots}
\usepackage{subfig}
\usepackage{xcolor}
\usepackage{multirow}
\usepackage{multicol}
\usetikzlibrary{plotmarks}
\usetikzlibrary{patterns}
\usepackage[bottom]{footmisc}

\interspeechcameraready 

\title{On the social bias of speech self-supervised models}

\name[affiliation={1}]{Yi-Cheng}{Lin}
\name[affiliation={1}]{Tzu-Quan}{Lin}
\name[affiliation={1}]{Hsi-Che}{Lin}
\name[affiliation={1,2}]{Andy T.}{Liu}
\name[affiliation={1}]{Hung-yi}{Lee}

\address{
  $^1$Graduate Institute of Communication Engineering, National Taiwan University , Taiwan\\
  $^2$ASUS Intelligent Cloud Services, Singapore}
\email{\{f12942075,hungyilee\}@ntu.edu.tw}


\begin{document}

\maketitle
\ninept
\setlength{\abovecaptionskip}{5pt}

\begin{abstract}

Self-supervised learning (SSL) speech models have achieved remarkable performance in various tasks, yet the biased outcomes, especially affecting marginalized groups, raise significant concerns. Social bias refers to the phenomenon where algorithms potentially amplify disparate properties between social groups present in the data used for training. Bias in SSL models can perpetuate injustice by automating discriminatory patterns and reinforcing inequitable systems. This work reveals that prevalent SSL models inadvertently acquire biased associations. We probe how various factors, such as model architecture, size, and training methodologies, influence the propagation of social bias within these models. Finally, we explore the efficacy of debiasing SSL models through regularization techniques, specifically via model compression. Our findings reveal that employing techniques such as row-pruning and training wider, shallower models can effectively mitigate social bias within SSL model.

\end{abstract}
\section{Introduction}
\label{sec:intro}
Self-supervised learning (SSL) models have proven outstanding performance in various tasks \cite{liu2021tera, wavlm, modified_cpc, vq_wav2vec, data2vec}, also called foundation models. The SSL framework includes two stages. The first stage pretrains a shared foundation model model on many unlabeled data with pretext objectives. In the second stage, the foundation models are adapted to downstream tasks supervised by extracting meaningful representations or fine-tuning the whole model. However, we observed SSL representation exhibiting bias and fairness problems that propagate from pretraining to downstream tasks, raising concerns about fairness. 

Machine learning models may inadvertently reinforce stereotypes by associating characteristics like emotion and hate more strongly with specific demographic attributes of the speaker, such as age, gender, or accent \cite{steed2021image}. For example, if a model consistently associates anger with young speakers and sadness with aged speakers, it could perpetuate stereotypes about emotional expression. Rectifying bias in SSL models poses considerable challenges due to the substantial economic and environmental costs involved \cite{NEURIPS2023_2ecc8008}. 


Most previous works evaluate social bias by the parity of downstream task performance between different social groups.
Comparison of performance differences between social groups for speech SSL models exists in Automated Speech Recognition
(ASR) \cite{boito2022study, dheram2022toward}, Speech Translation (ST) \cite{boito2022study}, and Emotion Recognition (ER) \cite{derington2023testing, wagner2023dawn}. 
Meng et al. \cite{meng2022dont} show that training data don't need to be gender balanced to reach the best downstream performance, and pretrain on data with lower speech rate has better downstream performance. There is also work on assessing classification accuracy difference for binary gender on ER models \cite{gorrostieta19_interspeech}. 
These works have studied downstream test performance differences, neglecting bias in the SSL model representation. Juliette et al. \cite{millet-dunbar-2022-self} demonstrated the self-supervised speech model's inclination towards the language it was trained on within the representation space, and SpEAT \cite{slaughter2023pretrained} revealed embedding-level bias between social groups in speech SSL models. However, whether the SSL features are more biased than the acoustic features is still unknown, and the contributing factors to social bias, such as model architecture and size, have not been discussed in previous works. 

While significant efforts have been devoted to improving the performance of SSL models, there has been a notable lack of research on the bias present in speech SSL models and the debiasing methods for social bias. 
Recognizing these research gaps for speech, we aim to answer the following research questions: 
(1) Do representations generated by SSL models amplify social bias? 
(2) Does model size, pertaining step, and architecture affect social bias? 
(3) Research in NLP shows a debiasing effect on compressed large language models \cite{gonçalves2023understanding, xu2022model, ramesh-etal-2023-comparative}. Do model compression techniques affect social bias in speech SSL models? 

To answer our research questions, we evaluate three well-known SSL models: HuBERT \cite{hsu2021hubert}, Wav2Vec2 \cite{baevski2020wav2vec}, and MelHuBERT \cite{lin2023melhubert}. 
To quantify the social bias of SSL model embedding in our research, we adopt SpEAT \cite{slaughter2023pretrained} to evaluate the gender, age, and nationality bias.

Compared to prior works, our research provides more solid evidence of SSL model's amplification of social bias and a more comprehensive exploration of various factors that might affect social bias. Notably, we pioneer the investigation into the efficacy of compression techniques in mitigating social bias within speech SSL models. The following discoveries and insights emerge from our investigation:
\begin{itemize}
    \item SSL model representation amplifies social bias compared to classic acoustic features such as spectrogram or Mel-Frequency Cepstral Coefficients (MFCC). 
    \item Model size has little effect on social bias. Also, among models with equal parameter counts, wider and shallower architectures typically exhibit reduced social bias compared to narrower and deeper models.
    \item Longer pretraining steps lead to higher social bias. Besides, models with identical training objectives and architectural designs exhibit consistent trends in bias during pretraining.
    \item Among the compression methods, row pruning offers effective mitigation strategies for all social bias in SSL speech models, and all compression methods can reduce \textit{Age} bias.
\end{itemize}

\section{Experiment setup}
\subsection{Speech Self-supervised models}
In recent years, speech self-supervised models have demonstrated strong capabilities in extracting general features \cite{mohamed2022self, morais2022speech}. Among those speech SSL models, Wav2Vec 2.0 \cite{baevski2020wav2vec} discretizes the latent feature space and learns contextualized representations through contrastive learning among masked time stamps. HuBERT's \cite{hsu2021hubert} loss function is inherited from wav2vec 2.0, but it moves the discretization process offline, using k-means clustering on either MFCC features or hidden representations to generate discrete targets. MelHuBERT \cite{lin2023melhubert} simplifies HuBERT's loss function to a simple cross-entropy loss and removes the convolutional layer, using Mel Spectrogram as the input directly, significantly reducing the computational cost of the model. 
Since these three SSL models have similar model architectures, they are easy to compare. In this paper, we will primarily focus on these three SSL models.

We train HuBERT, Wav2Vec2, and MelHuBERT models at smaller sizes in two different architectures: \textit{small} and \textit{slim}, with nearly the same number of parameters. Model details are shown in Table~\ref{tab:model-architecture}. 
To ensure alignment with \cite{hsu2021hubert}, HuBERT and Wav2Vec2 with configuration \textit{slim} and \textit{small} are trained using fairseq \cite{ott2019fairseq} with a maximum batch size of 87.5 seconds of audio per GPU across 400k steps, while for MelHuBERT we prepare with batch size 12 and 421k steps to ensure all models are trained with same amount of data instances. All models are trained with 960 hours of LibriSpeech \cite{librispeech} data.

We also investigate models with larger sizes, including HuBERT Base\footnote{https://dl.fbaipublicfiles.com/hubert/hubert\_base\_ls960.pt}, Wav2Vec2 Base\footnote{https://dl.fbaipublicfiles.com/fairseq/wav2vec/wav2vec\_small.pt}, Wav2Vec2 Large\footnote{https://dl.fbaipublicfiles.com/fairseq/wav2vec/libri960\_big.pt}, MelHuBERT-20ms \cite{lin2022compressing}. These models use the same dataset, LibriSpeech 960hr, for pretraining.

\subsection{Bias evaluation for Speech SSL models}
We use SpEAT to evaluate bias in Speech SSL models. SpEAT evaluates the alignment between two target concept embeddings, such as female ($X$) and male ($Y$), corresponding to its attribute concept embeddings, such as positive ($A$) and negative ($B$) valence. An effect size $d$ is calculated to show which of the target concepts $X$ or $Y$ is closer to attribute $A$ than to $B$. First, the difference of mean cosine similarity between a target embedding $w$ and each embedding $a\in A$, $b\in B$ is calculated:
\begin{align}
s(w,A,B) &= mean_{a\in A}cos(w,a)-mean_{b\in B}cos(w,b).%
\end{align}
And the effect size $d$ \cite{cohen2013statistical} is then calculated by:
\begin{align}
d &= \frac{mean_{x\in X} s(x,A,B)-mean_{y\in Y} s(y,A,B)}{std\_dev_{w\in X\cup Y} s(w,A,B)}.
\end{align}
SpEAT $d$ values larger than 0.20, 0.50, and 0.80 are viewed as small, medium, and large biases, respectively. Conversely, a negative $d$ value indicates a reverse bias.

We follow the SpEAT setup closely, employing attribute concepts of positive (A) and negative (B) valence. We turn to the Morgan Emotional Speech Set \cite{morgan2019categorical} to gather speech samples, following the setup in SpEAT. 
Our target concepts are drawn from the Speech Accent Archive \cite{weinberger2011speech}, with gender, age, and nationality filtered to create diverse bias categories. Table~\ref{tab:bias_category} provides an overview of these categories.
The embeddings for speech segments are calculated by averaging the embeddings within each layer along time dimension, followed by aggregating these averaged embeddings across all layers.
\begin{table}[hb!]
  \centering
  \caption{Bias categories in our experiment. \textbf{count} represents the number of stimuli per target group.}
  \vspace{-1.5mm}
  \begingroup
  \renewcommand*{\arraystretch}{0.8}
  \begin{tabular}{c c c c}
    \toprule
    \textbf{Category} & \textbf{target (X)} &  \textbf{target (Y)} & \textbf{count} \\
    \midrule
        \textit{Gender} & Female & Male & 60\\
        \textit{Native} & US & Korean & 55\\
        \textit{Age} & Young & Old & 58\\
    \bottomrule
  \end{tabular}
  \endgroup
  \vspace{-1mm}
  \label{tab:bias_category}
\end{table}
\colorlet{lightred}{red!40}
\colorlet{lightblue}{blue!40}
\colorlet{lightgreen}{green!40}

\pgfplotsset{
compat=1.11,
legend image code/.code={
\draw[mark repeat=2,mark phase=2]
plot coordinates {
(0cm,0cm)
(0.15cm,0cm)        
(0.3cm,0cm)         
};%
}
}

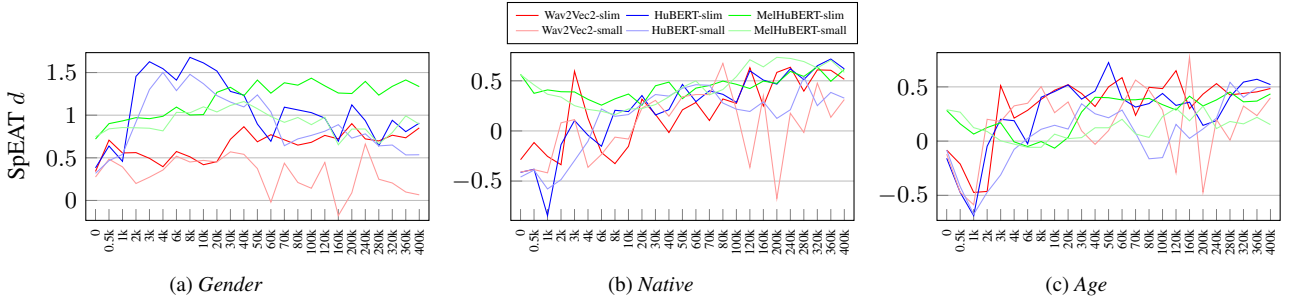
\begin{figure*}[h!]
    \centering
    \subfloat[Gender]{
        \begin{tikzpicture}
        \begin{axis}[  
        xtick pos=lower, ylabel=SpEAT $d$,
        ylabel near ticks,
        xticklabel style = {font=\tiny, rotate= 90},
        ytick pos=left, symbolic x coords={0,0.5k, 1k, 2k, 3k,4k,6k,8k,10k,20k,30k,40k,50k,60k,70k,80k,100k,120k,160k,200k,240k,280k,320k,360k,400k}, enlargelimits=0.03, xtick=data, grid=major, xmajorgrids=false, width=0.36\textwidth,
        height=3.8cm,]
        \addplot[mark=none, color=red] coordinates {(0, 0.34) (0.5k, 0.707) (1k, 0.557) (2k, 0.561) (3k, 0.494) (4k, 0.396) (6k, 0.574) (8k, 0.512) (10k, 0.419) (20k, 0.455) (30k, 0.717) (40k, 0.865) (50k, 0.688) (60k, 0.771) (70k, 0.711) (80k, 0.649) (100k, 0.682) (120k, 0.763) (160k, 0.722) (200k, 0.901) (240k, 0.722) (280k, 0.698) (320k, 0.765) (360k, 0.735) (400k, 0.851)};
        \addplot[mark=none, color=blue] coordinates {(0, 0.382) (0.5k, 0.638) (1k, 0.459) (2k, 1.455) (3k, 1.627) (4k, 1.545) (6k, 1.411) (8k, 1.678) (10k, 1.613) (20k, 1.518) (30k, 1.279) (40k, 1.235) (50k, 0.896) (60k, 0.694) (70k, 1.094) (80k, 1.062) (100k, 1.030) (120k, 0.966) (160k, 0.695) (200k, 1.121) (240k, 0.936) (280k, 0.642) (320k, 0.940) (360k, 0.807) (400k, 0.907)};
        \addplot[mark=none, color=green] coordinates {(0, 0.720) (0.5k, 0.901) (1k, 0.933) (2k, 0.971) (3k, 0.959) (4k, 0.987) (6k, 1.095) (8k, 1.002) (10k, 1.007) (20k, 1.268) (30k, 1.327) (40k, 1.228) (50k, 1.412) (60k, 1.258) (70k, 1.380) (80k, 1.352) (100k, 1.434) (120k, 1.344) (160k, 1.260) (200k, 1.254) (240k, 1.396) (280k, 1.235) (320k, 1.338) (360k, 1.413) (400k, 1.335)};
        \addplot+[mark=none, color=lightred, mark indices={ 2,3,6,7,8,9,10,11,12,13,15,18,21}] coordinates {(0, 0.275) (0.5k, 0.484) (1k, 0.394) (2k, 0.199) (3k, 0.273) (4k, 0.357) (6k, 0.519) (8k, 0.451) (10k, 0.470) (20k, 0.451) (30k, 0.570) (40k, 0.541) (50k, 0.374) (60k, -0.020) (70k, 0.437) (80k, 0.211) (100k, 0.143) (120k, 0.445) (160k, -0.178) (200k, 0.090) (240k, 0.661) (280k, 0.249) (320k, 0.205) (360k, 0.099) (400k, 0.066)};
        \addplot[mark=none, color=lightblue] coordinates {(0, 0.318) (0.5k, 0.467) (1k, 0.540) (2k, 0.920) (3k, 1.301) (4k, 1.502) (6k, 1.288) (8k, 1.479) (10k, 1.369) (20k, 1.230) (30k, 1.154) (40k, 1.097) (50k, 1.238) (60k, 1.039) (70k, 0.642) (80k, 0.722) (100k, 0.763) (120k, 0.812) (160k, 0.889) (200k, 0.731) (240k, 0.781) (280k, 0.640) (320k, 0.650) (360k, 0.534) (400k, 0.536)};
        \addplot[mark=none, color=lightgreen] coordinates {(0, 0.748) (0.5k, 0.840) (1k, 0.855) (2k, 0.852) (3k, 0.847) (4k, 0.816) (6k, 1.033) (8k, 1.028) (10k, 1.098) (20k, 1.038) (30k, 1.114) (40k, 1.160) (50k, 1.076) (60k, 0.984) (70k, 0.916) (80k, 0.973) (100k, 0.889) (120k, 0.976) (160k, 0.653) (200k, 0.837) (240k, 0.838) (280k, 0.665) (320k, 0.796) (360k, 0.992) (400k, 0.894)};
    
        \end{axis}
        \end{tikzpicture}
    }
    \subfloat[Native]{
        \begin{tikzpicture}
        \begin{axis}[  
        xtick pos=lower,
        ylabel near ticks,
        xticklabel style = {font=\tiny, rotate= 90},
        ytick pos=left, symbolic x coords={0,0.5k, 1k, 2k, 3k,4k,6k,8k,10k,20k,30k,40k,50k,60k,70k,80k,100k,120k,160k,200k,240k,280k,320k,360k,400k}, enlargelimits=0.03, xtick=data, grid=major,
        xmajorgrids=false, width=0.36\textwidth,
        legend style={at={(0.5,1.05)},anchor=south, nodes={scale=0.5, transform shape}},
        legend columns=3,
        height=3.8cm]
        \addplot[mark=none, color=red] coordinates {
        (0, -0.287) (0.5k, -0.116) (1k, -0.254) (2k, -0.337) (3k, 0.595) (4k, 0.126) (6k, -0.216) (8k, -0.326) (10k, -0.154) (20k, 0.316) (30k, 0.165) (40k, -0.016) (50k, 0.210) (60k, 0.284) (70k, 0.104) (80k, 0.319) (100k, 0.275) (120k, 0.630) (160k, 0.249) (200k, 0.584) (240k, 0.635) (280k, 0.397) (320k, 0.609) (360k, 0.605) (400k, 0.515)
        };
        \addplot[mark=none, color=blue] coordinates {
        (0, -0.413) (0.5k, -0.382) (1k, -0.845) (2k, -0.137) (3k, 0.104) (4k, -0.043) (6k, -0.153) (8k, 0.206) (10k, 0.193) (20k, 0.353) (30k, 0.159) (40k, 0.212) (50k, 0.464) (60k, 0.292) (70k, 0.400) (80k, 0.365) (100k, 0.285) (120k, 0.600) (160k, 0.507) (200k, 0.467) (240k, 0.620) (280k, 0.512) (320k, 0.650) (360k, 0.717) (400k, 0.619)
        };
        \addplot[mark=none, color=green] coordinates {
        (0, 0.564) (0.5k, 0.375) (1k, 0.409) (2k, 0.391) (3k, 0.389) (4k, 0.315) (6k, 0.257) (8k, 0.320) (10k, 0.368) (20k, 0.256) (30k, 0.450) (40k, 0.487) (50k, 0.322) (60k, 0.427) (70k, 0.453) (80k, 0.497) (100k, 0.465) (120k, 0.423) (160k, 0.497) (200k, 0.474) (240k, 0.594) (280k, 0.543) (320k, 0.639) (360k, 0.495) (400k, 0.614)
        };
        \addplot[mark=none, color=lightred] coordinates {
        (0, -0.410) (0.5k, -0.386) (1k, -0.418) (2k, 0.081) (3k, 0.113) (4k, -0.363) (6k, -0.234) (8k, -0.062) (10k, -0.081) (20k, 0.227) (30k, 0.305) (40k, 0.147) (50k, 0.341) (60k, 0.363) (70k, 0.363) (80k, 0.673) (100k, 0.209) (120k, -0.363) (160k, 0.350) (200k, -0.674) (240k, 0.176) (280k, -0.015) (320k, 0.476) (360k, 0.136) (400k, 0.313)
        };
        \addplot[mark=none, color=lightblue] coordinates {
        (0, -0.460) (0.5k, -0.391) (1k, -0.579) (2k, -0.486) (3k, -0.296) (4k, -0.103) (6k, 0.219) (8k, 0.146) (10k, 0.163) (20k, 0.268) (30k, 0.363) (40k, 0.347) (50k, 0.426) (60k, 0.461) (70k, 0.455) (80k, 0.278) (100k, 0.217) (120k, 0.194) (160k, 0.290) (200k, 0.125) (240k, 0.211) (280k, 0.538) (320k, 0.252) (360k, 0.384) (400k, 0.327)
        };
        \addplot[mark=none, color=lightgreen] coordinates {
        (0, 0.562) (0.5k, 0.459) (1k, 0.366) (2k, 0.338) (3k, 0.252) (4k, 0.215) (6k, 0.196) (8k, 0.162) (10k, 0.219) (20k, 0.223) (30k, 0.263) (40k, 0.340) (50k, 0.435) (60k, 0.499) (70k, 0.360) (80k, 0.413) (100k, 0.542) (120k, 0.710) (160k, 0.639) (200k, 0.734) (240k, 0.723) (280k, 0.692) (320k, 0.628) (360k, 0.706) (400k, 0.584)
        };

        \legend{Wav2Vec2-slim, HuBERT-slim, MelHuBERT-slim, Wav2Vec2-small, HuBERT-small, MelHuBERT-small}
        \end{axis}
        \end{tikzpicture}
    }
    \subfloat[Age]{
        \begin{tikzpicture}
        \begin{axis}[  
        xtick pos=lower,
        ylabel near ticks,
        xticklabel style = {font=\tiny, rotate= 90},
        ytick pos=left, symbolic x coords={0,0.5k, 1k, 2k, 3k,4k,6k,8k,10k,20k,30k,40k,50k,60k,70k,80k,100k,120k,160k,200k,240k,280k,320k,360k,400k}, enlargelimits=0.03, xtick=data, grid=major,
        xmajorgrids=false, width=0.36\textwidth,
        height=3.8cm]
        \addplot[mark=none, color=red] coordinates {(0, -0.083) (0.5k, -0.211) (1k, -0.473) (2k, -0.464) (3k, 0.511) (4k, 0.212) (6k, 0.284) (8k, 0.385) (10k, 0.469) (20k, 0.52) (30k, 0.439) (40k, 0.319) (50k, 0.5) (60k, 0.585) (70k, 0.238) (80k, 0.496) (100k, 0.484) (120k, 0.649) (160k, 0.299) (200k, 0.433) (240k, 0.53) (280k, 0.426) (320k, 0.44) (360k, 0.454) (400k, 0.485)};
        \addplot[mark=none, color=blue] coordinates { (0, -0.158) (0.5k, -0.473) (1k, -0.687) (2k, -0.047) (3k, 0.201) (4k, 0.188) (6k, -0.025) (8k, 0.404) (10k, 0.457) (20k, 0.518) (30k, 0.387) (40k, 0.462) (50k, 0.724) (60k, 0.392) (70k, 0.314) (80k, 0.346) (100k, 0.439) (120k, 0.329) (160k, 0.359) (200k, 0.145) (240k, 0.189) (280k, 0.408) (320k, 0.544) (360k, 0.57) (400k, 0.521)};
        \addplot[mark=none, color=green] coordinates {(0, 0.283) (0.5k, 0.159) (1k, 0.065) (2k, 0.125) (3k, 0.169) (4k, -0.007) (6k, -0.05) (8k, -0.004) (10k, -0.065) (20k, 0.034) (30k, 0.268) (40k, 0.404) (50k, 0.4) (60k, 0.38) (70k, 0.381) (80k, 0.394) (100k, 0.336) (120k, 0.287) (160k, 0.414) (200k, 0.324) (240k, 0.38) (280k, 0.452) (320k, 0.362) (360k, 0.369) (400k, 0.435)};
        \addplot[mark=none, color=lightred] coordinates {(0, -0.12) (0.5k, -0.472) (1k, -0.587) (2k, 0.201) (3k, 0.181) (4k, 0.325) (6k, 0.349) (8k, 0.502) (10k, 0.262) (20k, 0.361) (30k, 0.095) (40k, -0.03) (50k, 0.097) (60k, 0.333) (70k, 0.563) (80k, 0.469) (100k, 0.286) (120k, -0.29) (160k, 0.772) (200k, -0.471) (240k, 0.198) (280k, 0.009) (320k, 0.323) (360k, 0.235) (400k, 0.399)};
        \addplot[mark=none, color=lightblue] coordinates {(0, -0.083) (0.5k, -0.419) (1k, -0.675) (2k, -0.472) (3k, -0.312) (4k, -0.076) (6k, 0.025) (8k, 0.109) (10k, 0.146) (20k, 0.111) (30k, 0.344) (40k, 0.252) (50k, 0.215) (60k, 0.286) (70k, 0.075) (80k, -0.163) (100k, -0.152) (120k, 0.154) (160k, 0.025) (200k, 0.112) (240k, 0.212) (280k, 0.542) (320k, 0.404) (360k, 0.493) (400k, 0.497)};
        \addplot[mark=none, color=lightgreen] coordinates {(0, 0.289) (0.5k, 0.267) (1k, 0.124) (2k, 0.089) (3k, 0.003) (4k, -0.028) (6k, -0.057) (8k, -0.057) (10k, 0.064) (20k, 0.023) (30k, 0.032) (40k, 0.123) (50k, 0.123) (60k, 0.201) (70k, 0.066) (80k, 0.034) (100k, 0.227) (120k, 0.304) (160k, 0.187) (200k, 0.322) (240k, 0.116) (280k, 0.181) (320k, 0.159) (360k, 0.219) (400k, 0.152)};

        \end{axis}
        \end{tikzpicture}
    }
    \vspace{-0.2\baselineskip}
    \caption{Model training steps versus SpEAT effective size.}
    \label{fig:pretrain_step_pgf}
\end{figure*}

\subsection{Compression methods for Speech SSL models}
There are many different ways to compress speech SSL models. 
In this work, we follow Lin et al. \cite{lin2022compressing}, focusing on four basic compression methods: weight pruning, head pruning, row pruning, and knowledge distillation. 
Besides knowledge distillation, the other three methods can be considered iterative pruning. 
In iterative pruning, we prune the network with two steps iteratively. 
First, we prune blocks of weights and then perform finetuning on the remaining weights. 
Based on the definition of a block, pruning can be categorized into weight pruning, which is done at the level of individual weights; head pruning, which targets attention heads; and row pruning, which involves the rows and columns of the Feed-Forward Network (FFW).

\textbf{Pruning:} 
Among the three SSL models compared, HuBERT training has two stages; stage 2 requires the acoustic unit clustered by the latent representation of stage 1. Due to the unavailability of the stage-1 model and the resource constraints preventing us from training a new HuBERT model, we opt not to include HuBERT in the pruning experiments.

To perform head-pruning, we adopt a strategy of pruning the same number of heads per layer based on their L1 norm, as higher layers typically exhibit more significant L1 norms. This pruning occurs every 25k steps until 72 heads remain, followed by further pruning every 40k steps until only 12 heads remain. A learning rate of $5\times10^{-5}$ is employed throughout this process.

To implement row pruning, we target the rows of FFW1 and the columns of FFW2, with a pruning objective of the sum of their L1 norms. Specifically, we prune 128 dimensions every 25,000 steps until 512 rows remain by a learning rate of $10^{-5}$.

For weight pruning, we prune MelHuBERT first because the contrastive loss of Wav2Vec2 cannot indicate convergence of the model. We conduct iterative pruning on individual weights based on their L1 norm for all weights and biases in the linear layers of Transformers. We maintain an exponential moving average of the loss with a decay rate of 0.9998. Pruning is initiated if the loss remains within 0.001 compared to the loss 15,000 steps prior. Our pruning schedule varies in aggressiveness depending on the network's density: we prune 20\% until 20\% sparsity, 10\% until 50\% sparsity, 5\% until 65\% sparsity, 2.5\% until 70\% sparsity, 1\% until 80\% sparsity. We use a learning rate of $10^{-5}$. We recorded MelHuBERT's pruning schedule and applied it to Wav2Vec2.

All the pruned models are trained with Adam optimizer \cite{kingma2014adam}, with a batch size of 4 for MelHuBERT and 12 for Wav2Vec2, which aligns with previous works \cite{lin2022compressing, ott2019fairseq}.

\textbf{Distillation:} For distillation, we apply the same setup as DistilHubert \cite{chang2022distilhubert} in S3PRL \cite{mockingjay}. We distill the models to 2, 4, and 6-layer student networks, with an objective function to predict $4^{th}$ and $12^{th}$ layer of the original SSL model's representation. We use a learning rate of $2\times10^{-4}$ with Adam optimizer, batch size 24 for HuBERT, Wav2Vec2, and MelHuBERT. 
\begin{table}[ht]
\footnotesize
\centering
\setlength{\tabcolsep}{3pt}
\caption{Model details of MelHuBERT (Mel), HuBERT (HuB), and Wav2Vec2 (Wav) architectures. $n_l$ is the number of transformer layers. $d_m$ is the hidden size of encoder. $f_s$ is the dimensionality of the feed-forward layer. $n_h$ is the number of heads.}
\resizebox{\columnwidth}{!}{
    \begin{tabular}{c|cccc|ccc}
    \toprule
    \multirow{2}{*}{architecture}& \multicolumn{4}{c|}{configuration} &\multicolumn{3}{c}{parameters}\\
     & $n_l$ & $d_m$ & $f_s$ & $n_h$ & Mel & Wav & HuB \\
    \midrule
    Small & 3 & 640 & 2048 & 8 & 16.5M & 21.3M & 20.9M\\
    Slim & 12 & 384 & 768 & 8 & 15.6M & 20.4M & 20.0M\\
    Base & 12 & 768 & 3072 & 12 & 90.2M & 95.0M & 94.7M\\
    \bottomrule
    \end{tabular}
}
\label{tab:model-architecture}
\end{table}
\begin{figure}[h!]
    \captionsetup[subfigure]{justification=raggedleft,singlelinecheck=false,oneside,margin={0cm,0.65cm}}
    \hspace{1.3cm}
    \subfloat{\begin{tikzpicture} 
    \begin{axis}[
    height=1.6cm,
    hide axis,
    xmin=0,
    xmax=0.1,
    ymin=0,
    ymax=0.1,
    legend style={at={(0.5,1)},anchor=south,font=\scriptsize,nodes={scale=0.8, transform shape},column sep=4pt},
    legend columns=3
    ]
    \addlegendimage{blue}
    \addlegendentry{Gender}
    \addlegendimage{red}
    \addlegendentry{Native}
    \addlegendimage{green}
    \addlegendentry{Age}
    \addlegendimage{blue,dashed}
    \addlegendentry{Gender baseline}
    \addlegendimage{red,dashed}
    \addlegendentry{Native baseline}
    \addlegendimage{green,dashed}
    \addlegendentry{Age baseline}
    \end{axis}
\end{tikzpicture}}\vspace{-0.3cm}    \\
    \addtocounter{subfigure}{-1}
    \subfloat[Head pruning]{\begin{tikzpicture}
\begin{axis}[  
    width=0.58\columnwidth,
    height=3.5cm,
    enlargelimits=0.03,
    xlabel=count of remaining heads,
    xlabel style={font=\scriptsize, yshift=0.1cm},
    xtick pos=lower,
    xticklabel style={font=\tiny, rotate=90},
    xtick=data,
    xmajorgrids=false,
    symbolic x coords={144, 132, 120, 108, 96, 84, 72, 60, 48, 36, 24, 12},
    ylabel=SpEAT $d$,
    ylabel near ticks,
    ylabel style={font=\scriptsize,yshift=-0.15cm},
    ytick pos=left,
    yticklabel style = {font=\tiny},
]
\addplot[color=blue] coordinates {(132, 1.345) (120, 1.266) (108, 1.044) (96, 0.993) (84, 0.814) (72, 1.009) (60, 1.102) (48, 0.817) (36, 0.971) (24, 0.989) (12, 1.034)};
\addplot[color=red] coordinates { (132, 0.467) (120, 0.187) (108, 0.291) (96, 0.235) (84, 0.237) (72, 0.195) (60, 0.311) (48, 0.229) (36, 0.246) (24, 0.113) (12, 0.331)};
\addplot[color=green] coordinates {(132, 0.429) (120, 0.450) (108, 0.310) (96, 0.331) (84, 0.433) (72, 0.358) (60, 0.565) (48, 0.449) (36, 0.434) (24, 0.503) (12, 0.581)};
\addplot[dashed,color=blue] coordinates {(132, 1.210) (120, 1.210) (108, 1.210) (96, 1.210) (84, 1.210) (72, 1.210) (60, 1.210) (48, 1.210) (36, 1.210) (24, 1.210) (12, 1.210)};
\addplot[dashed,color=red] coordinates {(132, 0.209) (120, 0.209) (108, 0.209) (96, 0.209) (84, 0.209) (72, 0.209) (60, 0.209) (48, 0.209) (36, 0.209) (24, 0.209) (12, 0.209)};
\addplot[dashed,color=green] coordinates {(132, 0.506) (120, 0.506) (108, 0.506) (96, 0.506) (84, 0.506) (72, 0.506) (60, 0.506) (48, 0.506) (36, 0.506) (24, 0.506) (12, 0.506)};
\end{axis}
\end{tikzpicture}}
    \subfloat[Head pruning]{\begin{tikzpicture}
\begin{axis}[  
    width=0.58\columnwidth,
    height=3.5cm,
    enlargelimits=0.03,
    xlabel=count of remaining heads,
    xlabel style={font=\scriptsize, yshift=0.1cm},
    xtick pos=lower,
    xticklabel style = {font=\tiny, rotate=90},
    xtick=data,
    xmajorgrids=false,
    symbolic x coords={144, 132, 120, 108, 96, 84, 72, 60, 48, 36, 24, 12},
    ytick pos=left,
    yticklabel style = {font=\tiny},
]
\addplot[color=blue] coordinates { (132, 0.681) (120, 0.670) (108, 0.772) (96, 0.954) (84, 0.995) (72, 0.663) (60, 0.564) (48, 0.843) (36, 0.727) (24, 0.756) (12, 0.531)};
\addplot[color=red] coordinates {(132, 0.186) (120, 0.614) (108, 0.507) (96, -0.049) (84, 0.759) (72, 0.479) (60, 0.447) (48, 0.408) (36, 0.227) (24, 0.587) (12, 0.797)};
\addplot[color=green] coordinates {(132, 0.232) (120, 0.317) (108, 0.309) (96, -0.031) (84, 0.315) (72, 0.303) (60, 0.198) (48, -0.007) (36, -0.032) (24, 0.120) (12, 0.329)};
\addplot[dashed,color=blue] coordinates { (132, 0.730) (120, 0.730) (108, 0.730) (96, 0.730) (84, 0.730) (72, 0.730) (60, 0.730) (48, 0.730) (36, 0.730) (24, 0.730) (12, 0.730)};
\addplot[dashed,color=red] coordinates {(132, 0.436) (120, 0.436) (108, 0.436) (96, 0.436) (84, 0.436) (72, 0.436) (60, 0.436) (48, 0.436) (36, 0.436) (24, 0.436) (12, 0.436)};
\addplot[dashed,color=green] coordinates {(132, 0.333) (120, 0.333) (108, 0.333) (96, 0.333) (84, 0.333) (72, 0.333) (60, 0.333) (48, 0.333) (36, 0.333) (24, 0.333) (12, 0.333)};
\end{axis}
\end{tikzpicture}} 
    \vspace{-0.2cm}\\
    \subfloat[Row pruning]{\begin{tikzpicture}
\begin{axis}[  
    width=0.58\columnwidth,
    height=3.5cm,
    enlargelimits=0.03,
    xlabel=count of remaining rows,
    xlabel style={font=\scriptsize, yshift=0.1cm},
    xtick pos=lower,
    xticklabel style = {font=\tiny},
    xtick=data,
    xmajorgrids=false,
    symbolic x coords={2560, 2048, 1536, 1024, 512},
    ylabel=SpEAT $d$,
    ylabel style={font=\scriptsize,yshift=-0.15cm},
    ytick pos=left,
    yticklabel style = {font=\tiny},
]
\addplot[color=blue] coordinates {(2560, 1.017) (2048, 0.893) (1536, 0.971) (1024, 0.830) (512, 0.831)};
\addplot[color=red] coordinates {(2560, 0.212) (2048, -0.070) (1536, -0.093) (1024, 0.128) (512, 0.065)};
\addplot[color=green] coordinates {(2560, 0.443) (2048, 0.300) (1536, 0.304) (1024, 0.400) (512, 0.399)};
\addplot[dashed,color=blue] coordinates {(2560, 1.210) (2048, 1.210) (1536, 1.210) (1024, 1.210) (512, 1.210)};
\addplot[dashed,color=red] coordinates {(2560, 0.209) (2048, 0.209) (1536, 0.209) (1024, 0.209) (512, 0.209)};
\addplot[dashed,color=green] coordinates {(2560, 0.506) (2048, 0.506) (1536, 0.506) (1024, 0.506) (512, 0.506)};
\end{axis}
\end{tikzpicture}}        \hspace{-0.37cm}
    \subfloat[Row pruning]{\begin{tikzpicture}
\begin{axis}[  
    width=0.58\columnwidth,
    height=3.5cm,
    enlargelimits=0.03,
    xlabel=count of remaining rows,
    xlabel style={font=\scriptsize, yshift=0.1cm},
    xtick pos=lower,
    xticklabel style = {font=\tiny},
    xtick=data,
    xmajorgrids=false,
    symbolic x coords={2560, 2048, 1536, 1024, 512},
    ytick pos=left,
    yticklabel style = {font=\tiny},
]
\addplot[color=blue] coordinates {(2560, 0.167) (2048, 0.314) (1536, 0.404) (1024, 0.620) (512, 0.562)};
\addplot[color=red] coordinates {(2560, -0.561) (2048, -0.419) (1536, -0.515) (1024, 0.069) (512, 0.520)};
\addplot[color=green] coordinates {(2560, -0.503) (2048, -0.558) (1536, -0.786) (1024, -0.517) (512, -0.061)};
\addplot[dashed,color=blue] coordinates {(2560, 0.730) (2048, 0.730) (1536, 0.730) (1024, 0.730) (512, 0.730)};
\addplot[dashed,color=red] coordinates {(2560, 0.436) (2048, 0.436) (1536, 0.436) (1024, 0.436) (512, 0.436)};
\addplot[dashed,color=green] coordinates {(2560, 0.333) (2048, 0.333) (1536, 0.333) (1024, 0.333) (512, 0.333)};
\end{axis}
\end{tikzpicture}}     \vspace{-0.2cm}\\
    \subfloat[Weight pruning]{\begin{tikzpicture}
\begin{axis}[
    width=0.58\columnwidth,
    height=3.5cm,
    enlargelimits=0.03,
    xlabel=sparsity,
    xlabel style={font=\scriptsize, yshift=0.1cm},
    xtick pos=lower,
    xticklabel style = {font=\tiny},
    xtick=data,
    xmajorgrids=false,
    symbolic x coords={0, 0.2, 0.3, 0.4, 0.5, 0.6, 0.7, 0.8, 0.85, 0.9, 0.92, 0.94},
    ylabel=SpEAT $d$,
    ylabel near ticks,
    ylabel style={font=\scriptsize,yshift=-0.15cm},
    ytick pos=left,
    yticklabel style = {font=\tiny},
]
\addplot[color=blue] coordinates {(0, 1.210) (0.2, 1.290) (0.3, 1.211) (0.4, 1.167) (0.5, 1.051) (0.6, 1.232) (0.7, 1.124) (0.8, 1.102)};
\addplot[color=red] coordinates {(0, 0.209) (0.2, 0.252) (0.3, 0.346) (0.4, 0.434) (0.5, 0.555) (0.6, 0.382) (0.7, 0.475) (0.8, 0.468)};
\addplot[color=green] coordinates {(0, 0.506) (0.2, 0.577) (0.3, 0.486) (0.4, 0.512) (0.5, 0.462) (0.6, 0.351) (0.7, 0.466) (0.8, 0.460)};
\addplot[dashed,color=blue] coordinates {(0, 1.210) (0.2, 1.210) (0.3, 1.210) (0.4, 1.210) (0.5, 1.210) (0.6, 1.210) (0.7, 1.210) (0.8, 1.210)};
\addplot[dashed,color=red] coordinates {(0, 0.209) (0.2, 0.209) (0.3, 0.209) (0.4, 0.209) (0.5, 0.209) (0.6, 0.209) (0.7, 0.209) (0.8, 0.209)};
\addplot[dashed,color=green] coordinates {(0, 0.506) (0.2, 0.506) (0.3, 0.506) (0.4, 0.506) (0.5, 0.506) (0.6, 0.506) (0.7, 0.506) (0.8, 0.506)};
\end{axis}
\end{tikzpicture}}  \hspace{-0.13cm}
    \subfloat[Weight pruning]{\begin{tikzpicture}
\begin{axis}[
    width=0.58\columnwidth,
    height=3.5cm,
    enlargelimits=0.03,
    xlabel=sparsity,
    xlabel style={font=\scriptsize, yshift=0.1cm},
    xtick pos=lower,
    xticklabel style = {font=\tiny},
    xtick=data,
    xmajorgrids=false,
    symbolic x coords={0, 0.2, 0.3, 0.4, 0.5, 0.6, 0.7, 0.8},
    ytick pos=left,
    yticklabel style = {font=\tiny},
]
\addplot[color=blue] coordinates {(0, 0.730) (0.2, 0.760) (0.3, 0.571) (0.4, 0.830) (0.5, 0.912) (0.6, 0.728) (0.7, 0.751) (0.8, 0.576)};
\addplot[color=red] coordinates {(0, 0.436) (0.2, 0.215) (0.3, 0.024) (0.4, 0.130) (0.5, 0.451) (0.6, 0.531) (0.7, 0.622) (0.8, 0.903)};
\addplot[color=green] coordinates {(0, 0.333) (0.2, 0.272) (0.3, -0.200) (0.4, -0.001) (0.5, 0.107) (0.6, 0.055) (0.7, -0.016) (0.8, 0.420)};
\addplot[dashed,color=blue] coordinates {(0, 0.730) (0.2, 0.730) (0.3, 0.730) (0.4, 0.730) (0.5, 0.730) (0.6, 0.730) (0.7, 0.730) (0.8, 0.730)};
\addplot[dashed,color=red] coordinates {(0, 0.436) (0.2, 0.436) (0.3, 0.436) (0.4, 0.436) (0.5, 0.436) (0.6, 0.436) (0.7, 0.436) (0.8, 0.436)};
\addplot[dashed,color=green] coordinates {(0, 0.333) (0.2, 0.333) (0.3, 0.333) (0.4, 0.333) (0.5, 0.333) (0.6, 0.333) (0.7, 0.333) (0.8, 0.333)};
\end{axis}
\end{tikzpicture}}

    \caption{SpEAT $d$ versus parameters removed after applying 3 pruning methods on Wav2Vec2 (a, c, e) and MelHuBERT (b, d, f). The closer to the right of the graphs, the more parameters have been removed. The dashed lines are the SpEAT $d$ measured on the unpruned model.}
    \label{fig:pruning}
    \vspace{-1pt}
\end{figure}
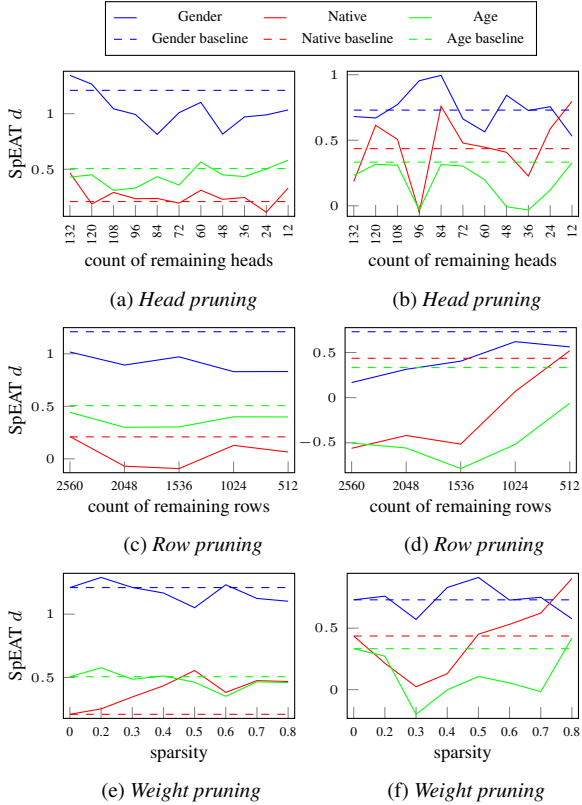

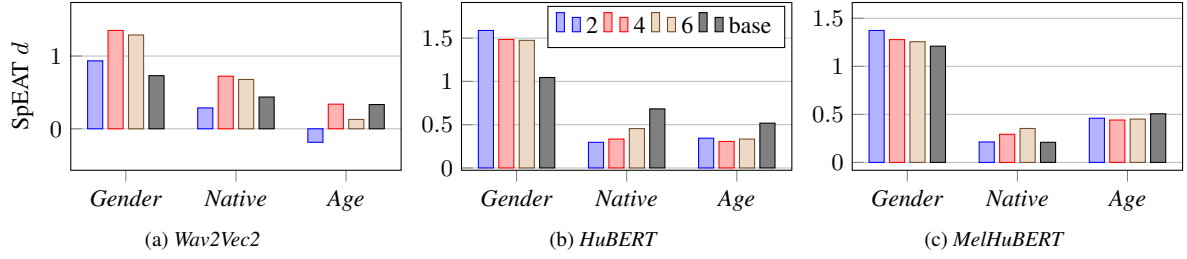
\begin{figure*}[h!]
    \centering
    \subfloat[Wav2Vec2]{
        \begin{tikzpicture}
        \begin{axis}[ 
        x tick label style={ /pgf/number format/1000 sep=}, 
        ybar, xtick pos=lower, ylabel=SpEAT $d$,
        ytick pos=left, symbolic x coords={\textit{Gender}, \textit{Native}, \textit{Age}}, enlargelimits=0.25, grid=major,
        xmajorgrids=false, bar width=0.2cm, width=0.35\textwidth,
        height=3.8cm,
        ]
        \addplot+ coordinates {(\textit{Gender}, 0.933927536)(\textit{Native}, 0.2870881259)(\textit{Age}, -0.185839057)};
        \addplot+ coordinates {(\textit{Gender}, 1.351376414) (\textit{Native}, 0.7229477763) (\textit{Age}, 0.3386466801)};
        \addplot+ coordinates {(\textit{Gender}, 1.28922379) (\textit{Native}, 0.6780483723) (\textit{Age}, 0.128529653)};
        \addplot+ coordinates {(\textit{Gender}, 0.7302286029) (\textit{Native}, 0.4364891052) (\textit{Age}, 0.3334083259)};
        \end{axis}
        \end{tikzpicture}
    }
    \subfloat[HuBERT]{
        \begin{tikzpicture}
        \begin{axis}[ 
        x tick label style={ /pgf/number format/1000 sep=}, 
        ybar, xtick pos=lower, 
        ytick pos=left, symbolic x coords={\textit{Gender}, \textit{Native}, \textit{Age}}, enlargelimits=0.25, grid=major,
        xmajorgrids=false, bar width=0.2cm, width=0.35\textwidth,
        legend columns=4,
        height=3.8cm,
        ]
        \addplot coordinates {(\textit{Gender}, 1.588111639) (\textit{Native}, 0.2962432504) (\textit{Age}, 0.3439335227)};
        \addplot coordinates {(\textit{Gender}, 1.483676672) (\textit{Native}, 0.3335299194) (\textit{Age}, 0.3062276542)};
        \addplot coordinates {(\textit{Gender}, 1.474191308) (\textit{Native}, 0.4547618926) (\textit{Age}, 0.3338756561)};
        \addplot coordinates {(\textit{Gender}, 1.044700146) (\textit{Native}, 0.6822436452) (\textit{Age}, 0.5174208879)};
        \legend{2, 4, 6, base}
        \end{axis}
        \end{tikzpicture}
    }
    \subfloat[MelHuBERT]{
        \begin{tikzpicture}
        \begin{axis}[ 
            x tick label style={/pgf/number format/1000 sep=}, 
            ybar,
            xtick pos=lower, 
            ytick pos=left, 
            symbolic x coords={\textit{Gender}, \textit{Native}, \textit{Age}},
            enlargelimits=0.25, 
            grid=major,
            xmajorgrids=false, 
            bar width=0.2cm, 
            width=0.35\textwidth, 
            height=3.8cm,
        ]
        \addplot coordinates {(\textit{Gender}, 1.373104095) (\textit{Native}, 0.2128697634) (\textit{Age}, 0.4599486589)};
        \addplot coordinates {(\textit{Gender}, 1.278155565) (\textit{Native}, 0.2928802669) (\textit{Age}, 0.441036582)};
        \addplot coordinates {(\textit{Gender}, 1.255961895) (\textit{Native}, 0.352996856) (\textit{Age}, 0.450789988)};
        \addplot coordinates {(\textit{Gender}, 1.209533691) (\textit{Native}, 0.2090277374) (\textit{Age}, 0.5060680509)};
        \end{axis}
        \end{tikzpicture}
    }        
    \vspace{-0.3\baselineskip}
    \caption{Effect of model distillation on (a) Wav2Vec2 (b) HuBERT (c) MelHuBERT. 2, 4, 6 stands for the number of layers in the distilled model. "base" stands for the model before distillation.}
    \label{fig:distillation_architercture}
\end{figure*}

\section{Result and Analysis}
\subsection{Effect of model architecture on bias}
Table~\ref{tab:model_size} illustrates that compared to traditional features, the features extracted by SSL models have a more significant effect size on three bias categories, indicating these models learned social bias from the training process. MelHuBERT, for instance, trained on Mel spectrograms, exemplifies increased bias. Notably, most SSL models have the highest SpEAT \textit{d} on \textit{Gender} across all bias categories.

Moreover, an increase in model size doesn't necessarily correlate with increased bias, contrasting with findings on compressing large language models \cite{gonçalves2023understanding}. For example, among the Wav2Vec2 series, \textit{large} model only has the greatest SpEAT \textit{d} on \textit{Native} bias. In contrast, the greatest SpEAT \textit{d} of \textit{Age} and \textit{Gender} appears in \textit{base} model. Additionally, although small models have comparable parameters to slim models, they tend to have smaller effect sizes overall, showing the effect of model architecture design on bias.

\subsection{Effect of training steps on bias}
\vspace{-2pt}
Figure~\ref{fig:pretrain_step_pgf} shows that models with the same training objective and architecture design show a similar trend in SpEAT $d$ during training. Moreover, biases tend to be more pronounced in the later stages of pretraining. Upon random initialization, MelHuBERT models consistently exhibit a significantly larger SpEAT $d$ than other models. Regardless of the initial SpEAT $d$, convergence to a positive value typically occurs within approximately 30k-60k steps. Among the three bias categories analyzed, \textit{Gender} emerges with the highest SpEAT $d$, suggesting a more significant embedding-level bias. 
\subsection{Effect of compression on bias}
To ensure the utility of compressed models, we first evaluate the upstream models on downstream task Phoneme Recognition (PR). MelHuBERT uses a much smaller batch size for pretraining, so the performance cannot be directly compared with Wav2Vec2 and HuBERT base \cite{vaessen2024effect}. We follow the evaluation configurations in the S3PRL toolkit. Table~\ref{tab:pruning_performance} shows all compressed models still exhibit reasonable performance on downstream tasks, compared with \cite{lin2022compressing}.
\begin{table}[htbp]
  \footnotesize
  \centering
  \caption{SpEAT effect size $d$ measured on speech SSL model representation or classical features with different sizes. The bold text denotes a model with less bias than any traditional features.}
  \vspace{-0.5mm}
  \begingroup
  \renewcommand*{\arraystretch}{0.88}
  \begin{tabular}{l|ccc}
    \toprule
    \textbf{Name} & \textbf{Age} &  \textbf{Native} & \textbf{Gender} \\
    \midrule
        Wav2Vec2 slim & 0.49 & 0.52 & 0.85 \\
        Wav2Vec2 small & 0.40 & 0.31 & \textbf{0.07} \\
        Wav2Vec2 base & 0.51 & 0.21 & 1.21 \\
        Wav2Vec2 large & 0.46 & 0.69 & 0.58 \\
        \midrule
        HuBERT slim & 0.52 & 0.62 & 0.91 \\
        HuBERT small & 0.50 & 0.33 & 0.54 \\
        HuBERT base & 0.52 & 0.68 & 1.04 \\
        \midrule
        MelHuBERT slim & 0.44 & 0.55 & 1.33 \\
        MelHuBERT small & \textbf{0.14} & 0.57 & 0.92 \\
        MelHuBERT base & \textbf{0.16} & 0.62 & 0.92 \\
        \midrule
        STFT & 0.39 & -0.07 & -0.04\\
        Spectrogram & -0.32 & -0.12 & 0.17\\
        Mel Spectrogram & -0.32 & -0.07 & 0.03\\
        MFCC & -0.05 & 0.13 & 0.50\\
    \bottomrule
  \end{tabular}
  \endgroup
  \label{tab:model_size}
\end{table}

\subsubsection{Head pruning}
We can see in Figure~\ref{fig:pruning}(a) and ~\ref{fig:pruning}(b) that head pruning reduces SpEAT $d$ in category \textit{Age}, showing an apparent debias effect. However, for \textit{Native} and \textit{Gender} bias, SpEAT $d$ fluctuates around the original model, suggesting a lack of significant debiasing effect in these aspects.

\subsubsection{Row pruning}
In Figure~\ref{fig:pruning}(c) and~\ref{fig:pruning}(d), we observe a general reduction in SpEAT $d$ across pruned models. Specifically, for the \textit{Gender} bias in MelHuBERT and for all bias categories in Wav2Vec2, there is a decreased bias. However, in the case of \textit{Native} and \textit{Age} bias in MelHuBERT, the SpEAT \textit{d} turns negative after pruning, indicating the anti-bias characteristics in these categories at the onset of pruning. Subsequently, SpEAT \textit{d} gradually increases as more parameters are pruned, leading to lower bias.

\subsubsection{Weight pruning}
Based on Figure~\ref{fig:pruning}(e) and ~\ref{fig:pruning}(f), it's evident that weight pruning generally lowers SpEAT $d$ in \textit{Age} but shows a varied result across other bias categories and models. Specifically for MelHuBERT, SpEAT $d$ initially decreases with increasing sparsity but then rises afterward.

\subsubsection{Distillation}
Figure~\ref{fig:distillation_architercture} vividly illustrates that model distillation does not universally alleviate bias but sometimes increases bias. 
In the case of distilled Wav2Vec2, we notice an increase in SpEAT $d$ for \textit{Gender} and \textit{Native} bias, yet a decrease for most models on \textit{Age} bias. 
Similarly, for distilled HuBERT, SpEAT $d$ escalates for \textit{Gender} but diminishes for \textit{Age} and \textit{Native}. 
Meanwhile, in the case of distilled MelHuBERT, there's an increase in SpEAT $d$ for \textit{Gender} while a decrease is observed for \textit{Native} and \textit{Age}. 
Distillation generally increases \textit{Gender} bias while decreases \textit{Age} bias. The distilled model size does not significantly affect the embedding level bias measured.

\begin{table}[htbp!]
\footnotesize
\setlength{\tabcolsep}{4pt}
\centering
\caption{Downstream performance of the compressed models. dist. represent distillation. For pruned models, we report the performance of models with middle sparsity (72 heads for head pruning, 1536 rows for row pruning, sparsity 0.4 for weight pruning) and the sparsest model, separated by "/". For distilled models, we report the performance of the 2-layered model. }
\begingroup
\renewcommand*{\arraystretch}{0.8}
\begin{tabular}{c|c|c|c}
\toprule
\multirow{2}{*}{\textbf{Model}} & \multirow{2}{*}{\textbf{Method}} & \multicolumn{2}{c}{\textbf{PR (PER$\downarrow$)}} \\
                           &   & base                 & compressed  \\
\midrule
\multirow{4}{*}{Wav2Vec2}  & dist.   & \multirow{4}{*}{5.7} & 21.3        \\
                           & head &  &
                           11.5 / 18.8 \\
                           & row  &  &
                           8.9 / 17.3  \\
                           & weight &  & 
                           6.6 / 8.9   \\
\midrule
HuBERT                     & dist.   & 5.41                 & 18.6        \\
\midrule
\multirow{4}{*}{MelHuBERT} & dist. & \multirow{4}{*}{8.2} &  27.2       \\
                           & head    &         & 7.0 / 11.6 \\
                           & row     &         & 8.8 / 14.1 \\
                           & weight  &        &  8.1 / 8.7 \\
\bottomrule
\end{tabular}
\endgroup
\label{tab:pruning_performance}
\end{table}

\section{Conclusion}
This study offers critical insights into the impact of various factors within speech SSL models on the embedding-level social bias. Our findings reveal that SSL representation amplifies social bias compared
to classic acoustic features. When holding parameter counts constant, the \textit{small} model exhibits lower bias than the \textit{slim} model. Longer pretraining steps lead to higher social bias. Furthermore, row pruning demonstrates a capacity to reduce social bias among the compression techniques examined, whereas weight pruning, head pruning, and distillation show limited effectiveness. Lastly, all these pruning methods effectively decrease \textit{Age} bias in upstream models. Our research serves as a roadmap toward developing more equitable and efficient SSL models.

\section{Limitation}
The analysis in our work focuses primarily on speaker-specific traits such as $Gender$, $Native$, and $Age$, potentially overlooking other dimensions of bias that could exist in semantic components of speech data such as political ideology and religion. Also, our analysis focuses on models trained with mainly English datasets. Further research is required on whether these results can apply to other languages or multilingual settings.

\section{Acknowledgement}
We thank to National Center for High-performance Computing (NCHC) of National Applied Research Laboratories (NARLabs) in Taiwan for providing computational and storage resources.

\bibliographystyle{IEEEtran}
\bibliography{mybib}

\end{document}